\documentclass[preprint,notoc]{JHEP} 
\usepackage{amsmath}
\usepackage{amssymb,amsfonts}
\usepackage{epsfig}

\newcommand\fverb{\setbox\pippobox=\hbox\bgroup\verb}
\newcommand\fverbdo{\egroup\medskip\noindent%
            \fbox{\unhbox\pippobox}\ }
\newcommand\fverbit{\egroup\item[\fbox{\unhbox\pippobox}]}
\newbox\pippobox

\newcommand{\be}{\begin{equation}}
\newcommand{\ee}{\end{equation}}
\newcommand{\ben}{\begin{enumerate}}
\newcommand{\een}{\end{enumerate}}
\renewcommand{\sp}{\ ,\qquad}
\makeatletter
\renewcommand{\@makefnmark}{\mbox{$^{\ddagger\@thefnmark}$}}
\renewcommand{\subsection}{\@startsection
  {subsection}{2}{0pt
}{-\baselineskip}{0.5\baselineskip}
  {\normalfont\normalsize\itshape}}
\makeatother
\renewcommand{\theequation}{\arabic{equation}}
\numberwithin{table}{section}

\newcommand{\Ord}{{\cal{O}}}

\title{Hagedorn Behaviour of Little String Theory from String Corrections
to NS5-Branes}

\author{T. Harmark and N.A. Obers\thanks{
Work supported in part by TMR network ERBFMRXCT96-0045.}\\
    Niels Bohr Institute and Nordita, Blegdamsvej 17, DK-2100 Copenhagen, Denmark \\
    E-mail: \email{harmark@nbi.dk}, \email{obers@nordita.dk} }

\preprint{ NBI-HE-00-22  \\
NORDITA-2000/39 HE \\
\hepth{0005021}}    

\abstract{ Following the conjectured duality between near-horizon
NS5-branes and little string theory, the string-corrected
thermodynamics of near-horizon NS5-branes is studied and found to
agree with the statistical thermodynamics of a 5+1 dimensional
supersymmetric string theory near the Hagedorn temperature.
Specifically, tree-level corrections to the temperature are argued
to vanish, in accordance with the duality, while the one-loop
string correction to the NS5-brane thermodynamics is shown to
generate the correct temperature dependence of the entropy. }

\begin{document}

\maketitle 

\newcommand{\bhg}{\beta_{\rm hg}}
\newcommand{\thg}{T_{\rm hg}}

\section{Introduction}
The near-horizon limit of NS5-branes is conjectured to be dual
to Little String Theory (LST) with 16 supercharges \cite{Aharony:1998ub}.
LST is a 5+1 dimensional non-gravitational and
non-local theory of strings
\cite{Seiberg:1997zk}%
\footnote{See also Refs. \cite{Dijkgraaf:1997ku,Losev:1997hx} and
\cite{Aharony:1999ks} for a brief review of LST.}. As for any
string theory, the statistical mechanics description of LST breaks
down at a certain temperature, known as the Hagedorn temperature.
This raises the question whether it is possible to observe, via
the conjectured near-horizon-NS5/LST duality, Hagedorn behaviour
of LST from the thermodynamics of near-horizon NS5-branes.
Previous work on the subject
\cite{Maldacena:1996ya,Maldacena:1997cg} has revealed that the
leading order thermodynamics of non-extremal near-horizon
NS5-branes indeed corresponds to the leading order Hagedorn
behaviour, since one finds the relation $ E = T S $ where $T$ is
constant. Thus, the seemingly peculiar behaviour of the
thermodynamics of near-horizon NS5-branes is exactly the behaviour
needed in order for the near-horizon-NS5/LST duality to work.

In this letter we go a step further by including string
corrections in the thermodynamics of near-horizon NS5-branes. The
string-corrected thermodynamics for near-horizon NS5-branes is
particularly interesting since the leading order thermodynamics
has a constant temperature, so that the thermodynamical phase space
is completely degenerate%
\footnote{In \cite{Harmark:1999xt} it has been shown that the
thermodynamics of near-horizon spinning 5-branes in type II string
theory has a degenerate phase space as well, {\it e.g.} $T^2 + \Omega^2
= {\rm const.}$ for one angular momentum. The near-horizon limit
of Reissner-Nordstr{\"o}m black holes also leads to degenerate
thermodynamics, though in a different way
\cite{Kiritsis:1999ke}.}. Corrections to the temperature, however,
expand the phase space and make small deviations from the
(constant) leading order temperature possible. Studying the
string-corrected thermodynamics therefore allows the determination
of the temperature dependence of the leading order entropy and
energy. In this sense, the string corrections to the temperature
that we find are not subleading since they are part of the
leading order thermodynamic quantities when expressed in terms of
the temperature.

The  thermodynamics obtained is compared to the statistical
thermodynamics of a 5+1 dimensional supersymmetric closed string
theory (reviewed in the appendix) and agreement is found.
In particular, we reproduce the entropy 
\begin{equation}
\label{entr0} S(T) = k \frac{\thg}{\thg- T} \ , 
\end{equation} 
where $k$ is a constant determined by the  one-loop corrected
near-horizon NS5-brane background. Equivalently, the entropy \eqref{entr0}
can be written in the form  
\begin{equation}
S(E) =\bhg E + k \log E \ . 
\end{equation}
This is therefore an example of agreement between statistical thermodynamics
of a non-gravitational theory on the one hand and Bekenstein-Hawking
thermodynamics of a black brane on the other hand.
We note that we can compute the thermodynamics from both sides
since we are in a regime in which both
LST and the bulk string theory description are weakly coupled.
Our successful comparison lends further confidence to the conjecture
that the various dualities between string and M-theory on
near-horizon brane backgrounds
and certain non-gravitational theories hold
beyond the leading order supergravity solution.

The non-trivial agreement we find between string-corrected macroscopic
Bekenstein-Hawking thermodynamics and thermodynamics derived
from microscopic degrees of freedom extends the
microscopic derivations of black hole entropies,
initiated in \cite{Strominger:1996sh}.
Our comparison, however, is different from \cite{Strominger:1996sh}, 
since no extrapolation from weak to
strong coupling needs to be made.

\section{One-loop corrected thermodynamics of NS5-branes}
\setcounter{equation}{2}

Six-dimensional LSTs with 16 supercharges can be defined
from $N$ coincident NS5-branes
in the limit \cite{Seiberg:1997zk}
\begin{equation}
\label{decoup} g_s \rightarrow 0 \sp l_s = \mbox{fixed} \ ,
\end{equation}
with $g_s$ and $l_s$ being respectively the string coupling
and string length of type II string theory.
In the limit \eqref{decoup} the bulk modes decouple and
give rise to non-gravitational theories: The $(1,1)$ LST for type IIB
NS5-branes and the $(2,0)$ LST for type IIA NS5-branes, both
of type  $A_{N-1}$ \footnote{It is also possible to
have LSTs of type $D$ and $E$ \cite{Diaconescu:1997gu,Aharony:1998ub}.}.

Moreover, it is believed that string theory in the background of
$N$ NS5-branes is dual to LST \cite{Aharony:1998ub}, in parallel
with the AdS/CFT correspondence \cite{Maldacena:1997re}. To this
end, consider the supergravity solution of $N$ coincident
NS5-branes in type II string theory with $r$ the transverse radius
and $r_0$ the horizon radius. Then, if we take the decoupling
limit \eqref{decoup} keeping fixed the energy scales
\begin{equation}
\label{ufixed}
u = \frac{r}{g_s l_s^2} \sp
u_0 = \frac{r_0}{g_s l_s^2}
\end{equation}
one obtains the Einstein-frame metric
\begin{equation}
\label{NS5met}
\frac{ds^2}{\sqrt{g_s} l_s}
 = \sqrt{\frac{u}{\sqrt{N} l_s }}
 \left[ - \left(1-\frac{u_0^2}{u^{2}} \right) dt^2
+ \sum_{i=1}^5 (dy^i)^2 + N l_s^2 \left( \frac{du^2}{u^2-u_0^2} +
d\Omega_{3}^2 \right) \right] \ ,
\end{equation}
and the dilaton
\begin{equation}
\label{NS5dil} g_s e^\phi = \frac{\sqrt{N}}{l_s u} \ .
\end{equation}
This supergravity solution is conjectured to be dual to a LST
\cite{Aharony:1998ub}.
The variable $u$ defined by \eqref{ufixed} is kept finite
since in type IIB it corresponds  to the mass
of an open D-string stretching between two NS5-branes with
distance $r$. In type IIA, $u/l_s$  is instead the induced
string tension of an open D2-brane
stretching between two NS5-branes.

The curvature $e^{-\phi/2}R$ in units of $l_s$ is of order
\begin{equation}
\label{epsD} \varepsilon_D = \frac{1}{N} \ ,
\end{equation}
while the effective string coupling squared at the horizon radius
is of order
\begin{equation}
\label{epsL} \varepsilon_L =  \frac{N}{l_s^2 u_0^2} \ .
\end{equation}
In order for the near-horizon NS5-brane solution \eqref{NS5met},
\eqref{NS5dil} to be a dual description of LST we need that
\( \varepsilon_D \ll 1 \) and $\varepsilon_L \ll 1$
 which is fulfilled for \cite{Maldacena:1997cg,Aharony:1998ub}
\begin{equation}
\label{limits} N \gg 1 \sp l_s^2 u_0^2 \gg N \ ,
\end{equation}
so that near-horizon NS5-branes describe LST in the UV-region.

The thermodynamics%
\footnote{The thermodynamics of spinning near-horizon NS5-branes is
found in \cite{Sfetsos:1999pq,Harmark:1999xt}.}
of the near-horizon NS5-brane solution is
\cite{Maldacena:1997cg}
\begin{equation}
\label{leadtherm}
T = \frac{1}{2\pi \sqrt{N} l_s} \sp
S = \frac{\sqrt{N} V_5}{(2\pi)^4 l_s^3 } u_0^2 \sp
E = \frac{V_5}{(2\pi)^5 l_s^4} u_0^2 \sp
F = 0 \ .
\end{equation}
As previously stated \cite{Maldacena:1996ya,Maldacena:1997cg},
this thermodynamics can be seen as the zeroth order approximation
to the thermodynamics of a string theory at high temperature. By
identifying the Hagedorn temperature as
\begin{equation}
\label{thag}
\thg = \frac{1}{2\pi \sqrt{N} l_s}
\end{equation}
the entropy takes the form
\begin{equation}
S(E) = \bhg E \ ,
\end{equation}
which is the leading order part of \eqref{hagentr}.
{}From the formula \eqref{canthg} we see that for a 5+1 dimensional
supersymmetric string theory the Hagedorn temperature is
\begin{equation}
\thg = \frac{1}{2\pi \hat{l}_s} \ ,
\end{equation}
where $\hat{l}_s$ is the string length of the theory.
Comparing with \eqref{thag}, it follows that
\begin{equation}
\label{strlen}
\hat{l}_s = \sqrt{N} l_s
\end{equation}
is the string length associated with the Hagedorn exponential growth of
string states in LST of type $A_{N-1}$.

If we write $\tau = (2\pi l_s^2)^{-1}$ as the ordinary string tension
and $\hat{\tau} = (2\pi \hat{l}_s^2)^{-1}$
as the string tension associated with the Hagedorn behaviour, we have
\begin{equation}
\label{strten} \hat{\tau} = \frac{1}{N} \tau \ .
\end{equation}
This shows that the string tension associated with the Hagedorn behaviour
is quantized in a unit that is a fraction of
the ordinary string tension.
This is in fact a well known phenomena.
If we consider LST as the decoupling limit \eqref{decoup}
of type II string theory on an $A_{N-1}$ singularity \cite{Ooguri:1996wj},
it is known from the study of string theory on this
orbifold singularity \cite{Douglas:1996sw} that
there exists so-called fractional branes which have tensions $1/N$
relative to the same type of brane in ten dimensions%
\footnote{We note that T-duality of LST with 16 supercharges together
with S-duality of $(2,0)$ LST \cite{Seiberg:1997zk,Losev:1997hx} connects
all the fractional little branes.}.
Thus, the string tension \eqref{strten} is the tension
of a fractional fundamental string in LST.
The fractional branes are in a sense more fundamental since one can
construct branes with tension being multiples of the fractional brane
tension by stacking them on top of each other.
Fractional strings can also be seen \cite{Hashimoto:1996pd}  by studying
the D1-D5 bound state which provides an alternative description of 
strings in six dimensions.

In summary, there are two energy scales associated with strings
appearing in LST of type $A_{N-1}$, the ordinary string scale $M_s
= l_s^{-1}$ and the fractional string scale $\hat{M}_s = M_s /
\sqrt{N}$. Clearly, the Hagedorn temperature picks the $\hat{M}_s$
string scale since this is the first one reached when raising the
temperature \cite{Aharony:1998ub}. Thus the Hagedorn
temperature of a string theory exhibits in this way the smallest 
quantized unit of string charge.

We note that the Hagedorn exponential growth of states in LST can
also be described by strings with string length $l_s$, but in that
case the sigma-model will have central charge $c=6N$,  as can be
seen by  comparing \eqref{thag} with \eqref{canthg}
\cite{Maldacena:1996ya}.

We now want to include string corrections in the thermodynamics%
\footnote{We refer to \cite{Correia:2000} for computational
details and a general analysis of string corrections to near-horizon
brane thermodynamics.}.
String theory has two expansions, the derivative expansion
and the string loop expansion.
It follows from \eqref{epsD} that the derivative expansion,
which is an expansion in $\alpha'=l_s^2$,
becomes an expansion in $\varepsilon_D$ for the
near-horizon NS5-brane solution.
The string loop expansion becomes instead an expansion in
$\varepsilon_L$ given in \eqref{epsL}.
By including these corrections we are thus testing the
conjecture of the near-horizon-NS5/LST correspondence beyond
the leading order supergravity solution.

In type II string theory the first corrections in the derivative
expansion appear at order $\varepsilon_D^3$, generated by the
well-known $R^4$ coupling \cite{Gross:1986iv} and its
supersymmetric completion. Organizing these terms in the string
loop expansion, there is a tree-level term at order
$\varepsilon_D^3$ and a one-loop term at order $\varepsilon_D^3
\varepsilon_L$. In type IIB there are also non-perturbative
contributions \cite{Green:1997tv}, but we will not consider these
here since they are subleading.

The tree-level expansion of the temperature is
\begin{equation}
\label{Ttree} T = \frac{1}{2\pi \sqrt{N} l_s} \left( 1 +
\sum_{i=3}^\infty a_i \frac{1}{N^i} \right) \ .
\end{equation}
This seems to indicate that the Hagedorn temperature is given by
\eqref{thag} only when $N$ is large whereas for finite $N$ it has a
different $N$-dependence.
Nevertheless, we expect that there are no tree-level corrections
to the Hagedorn temperature so that \( a_i = 0 \) for all $i$.
This is easily seen to be necessary from the requirement
that \eqref{Ttree}  be equal to \eqref{canthg}, since from \eqref{Ttree}
it would follow that the string tension $\hat{\tau}$
associated with Hagedorn behaviour of LST
would have corrections $\hat{\tau} = N^{-1} \tau ( 1 + \Ord(N^{-1}) )$
which clearly cannot make sense as a fractional string tension.

Also, if we take the other point of view and keep $\tau$ as the
tension of the strings associated with the Hagedorn behaviour of LST,
then \eqref{Ttree} can only be equal to
\eqref{canthg} if the central charge receives corrections
of the form $c = 6N ( 1 + \Ord(N^{-1}) ) $.
This is clearly impossible, {\it e.g.} it has been shown in the CHS model
that the central charge has to be a multiple
of $6$ \cite{Callan:1991dj}.

Thus, in order for the near-horizon-NS5/LST correspondence to work, it
is necessary that the tree-level corrections to the temperature
of the near-horizon black NS5-branes vanish.
Actually, this fact can be provided by noticing that the near-horizon
black NS5-brane solution is described by the
$SL(2,\mathbb{R})/SO(1,1) \times SU(2)$ exact CFT
\cite{Maldacena:1997cg,Sfetsos:1998wc}.
This implies that the metric in the string frame
does not have tree-level corrections \cite{Tseytlin:1993df},
which in turn leads to the result that the temperature has no
tree-level corrections.

The leading correction to the temperature is therefore generated by
the one-loop term of order
$\varepsilon_D^3 \varepsilon_L$, discussed above.
Thus, using \eqref{epsD}, \eqref{epsL} we can write
\begin{equation}
\label{Tcorr} T = \thg \left( 1 - \frac{\pi^2 b}{2} \frac{1}{N^2
l_s^2 u_0^2} \right) \ ,
\end{equation}
where $\thg$ is given by \eqref{thag} and the constant $b$ is a
rational number which depends on the exact form of the perturbed
solution. Since we assume that all tree-level corrections to the
temperature vanish, \eqref{Tcorr} is valid for all \( u_0 \gg
\sqrt{N} l_s^{-1} \). As a consequence of \eqref{Tcorr}, the
thermodynamics of the string-corrected NS5-brane solution is
qualitatively very different from the uncorrected one. Without
corrections the temperature is constant which means that the
thermodynamical phase space is degenerate. Including the string
corrections on the other hand, the temperature will depend on
$u_0$ as for any other $p$-brane in string and M-theory. As we
shall see, the fact that the temperature only depends on $u_0$
when the string loop corrections are included means that the
corrected temperature \eqref{Tcorr} determines the leading order
behaviour of the thermodynamics. On general grounds, it is
expected that the temperature $T$ in \eqref{Tcorr} is an
increasing function of the energy, which means that $b > 0$ in
\eqref{Tcorr}. Assuming $b > 0$ we see that \eqref{Tcorr} strongly
suggest that the temperature $\thg$ is a limiting Hagedorn
temperature of LST, in the sense that the temperature $T$ of the
system can come arbitrarily close to $\thg$ for higher and higher
energy $E \propto u_0^2$, but that it never can pass $\thg$.

Solving \eqref{Tcorr} for $u_0^2$ we obtain from \eqref{leadtherm}
the entropy
\begin{equation}
\label{ST} S(T) = \frac{b V_5}{32 \pi^2 N^{3/2} l_s^{5}}
\frac{\thg}{\thg-T} \ .
\end{equation}
Using the fractional string length \eqref{strlen} this can be written
\begin{equation}
\label{SofT} S(T) = \pi^3 b N \hat{V}_5 \frac{\thg}{\thg - T} \ ,
\end{equation}
where we have defined
\begin{equation}
\hat{V}_5 = \frac{V_5}{(2\pi \hat{l}_s)^5 } \ .
\end{equation}
{}From \eqref{SofT} we then obtain the free energy and energy
\begin{equation}
F(T) =  \pi^3 b N \hat{V}_5 \log \left( \frac{\thg - T}{\thg}
\right) \ ,
\end{equation}
\begin{equation}
\label{EofT} E(T) = \pi^3 b N \hat{V}_5  \thg \left[ \frac{T}{\thg
- T} + \log \left( \frac{\thg - T}{\thg} \right) \right] \ .
\end{equation}
Eqs. \eqref{ST}-\eqref{EofT} are valid for temperatures satisfying
\begin{equation}
\label{Trange} \frac{\thg- T}{\thg} \ll \frac{1}{N^3} \ .
\end{equation}
so that the temperature phase space,
which is degenerate at leading order, has now acquired a non-zero
width. Finally, \eqref{SofT} and \eqref{EofT} determine the
entropy as a function of energy
\begin{equation}
\label{SE} S(E) = \bhg E + \pi^3 b N \hat{V}_5 \log( E ) \ .
\end{equation}

We are now in position to compare with the statistical
thermodynamics of a 5+1 dimensional closed supersymmetric string theory.
To this end we note that
from \eqref{leadtherm} and \eqref{limits}
we have
\begin{equation}
\hat{l}_s E = \hat{V}_5 N^3 l_s^2 u_0^2 \gg N^4 \hat{V}_5 \ ,
\end{equation}
which clearly implies that
\begin{equation}
\hat{l}_s E \gg ( \hat{V}_5 )^{1/5} \ .
\end{equation}
The 5 spatial world-volume dimensions are thus effectively compact
in the energy range where LST is described
by near-horizon NS5-branes.
Therefore one should compare the string-corrected thermodynamics 
\eqref{SofT} and \eqref{SE}
with the one obtained for closed strings
living in 5 compact dimensions, which corresponds to $d=0$ in the appendix.

We then find that the entropy \eqref{SofT} of the one-loop corrected NS5-brane
solution and the entropy \eqref{hagentrT} derived by statistical mechanics
for closed strings agree in that they exhibit identical temperature dependence. 
This means that the constant of proportionality in \eqref{hagentrT} for
LST with 16 supercharges is predicted to be 
\begin{equation}
\label{k} k = \pi^3 b N \hat{V}_5 \ .
\end{equation}
It should be noted that for the matching of the string-corrected
near-horizon thermodynamics with the statistical thermodynamics,
it is of crucial importance that (i) we have shown that $d=0$ and
(ii) LST is a closed string theory. The reason being that if $d\neq
0$ or if there is an open string sector, then the statistical
thermodynamics is completely different
\cite{Salomonson:1986eq,Lowe:1995nm} from that of \eqref{hagentr}.

Moreover, it is important for the matching that we can argue that
LST is weakly coupled for the energies under consideration, since
the statistical thermodynamics in the appendix is that of a weakly
coupled string gas at high temperatures. By comparing
\cite{Losev:1997hx} the tension of induced little d-branes with
that of the corresponding open D-branes stretching between two
NS5-branes a distance $r_0$ apart, we find that the induced string
coupling%
\footnote{We note that the induced string coupling \( g_{\rm LST}
\) is a natural definition of the coupling since {\it e.g.} for the
$(2,0)$ LST it is equal to \( \tau / \tau_{\rm d1} \) where \(
\tau_{\rm d1} \) is the tension of the little d1-brane. Thus it
measures the mass scale of the perturbative modes relative to the
non-perturbative modes, and in the limit \( g_{\rm LST} \rightarrow
0 \) only the fundamental string modes are dynamical. The string
coupling \( g_{\rm LST} = (u_0 l_s)^{-1} \) has also been obtained
in the so-called Double-Scaling limit \cite{Giveon:1999px}.} in
LST is \( g_{\rm LST} = (u_0 l_s)^{-1} \). It then follows from
\eqref{limits} that in our range of energies the strings are
weakly coupled.

{}From \eqref{k} we see that $k \gg 1$. Therefore, as stated in the
appendix, the Canonical and Microcanonical ensembles
are equivalent.
In previous work on the Hagedorn behaviour of string theory
it has been found that the Canonical and Microcanonical ensembles
are inequivalent \cite{Brandenberger:1989aj,Lowe:1995nm}.
The reason for this inequivalence of ensembles
in ordinary string theory is that for sufficiently high
energy densities the Jeans instability of a gravitational system
sets in and the energy
fluctuations become too large for the Canonical ensemble to be
well-defined \cite{Brandenberger:1989aj,Lowe:1995nm}.
In our case, however, we have a string theory without gravity%
\footnote{See Refs. \cite{Barbon:1998cr} 
for recent work on Hagedorn behaviour in theories with gravity.}   
and thus the Jeans instability cannot be present.
Therefore, the equivalence between ensembles
is precisely in accord with the properties of LST.

As discussed in \cite{Salomonson:1986eq,Lowe:1995nm} the general
phase structure of a closed string theory with single string
density of levels \( E^{-a} \exp( \bhg E ) \) includes
a so-called long string phase, when the
energy density reaches a certain critical energy density.
For $a > 1$, the system is dominated in this phase by one single long string
that carries all the extra energy density above the critical
energy density. This phase has therefore negative heat capacity.
For $a=1$, which corresponds to LST for the energies we consider,
the long string phase consists instead of a condensate of several
long strings which carry all the extra energy density above the
critical energy density. In this case the heat capacity is
positive since the long string condensate spreads out the extra
energy uniformly in the system. This uniform distribution of
energy is accomplished by the fact that the long strings can split
and join. In our system, we find from \eqref{SofT} the specific heat
\begin{equation}
C(T) = \pi^3 b N \hat V_5 \frac{\thg T}{(\thg-T)^2} \ ,
\end{equation}
which is manifestly positive.

It is not difficult to include the effects of higher-loop
corrections
 in the analysis above. Using \eqref{epsL}, the temperature takes the form,
\begin{equation}
\label{Thloop} T = \thg \left[ 1 - \frac{\pi^2 b}{2} \frac{1}{N^2
l_s^2 u_0^2} + \sum_{g=2}^\infty \frac{b_g}{N^{d_g}} \left(\frac{N}{l_s^2
u_0^2}\right)^g \right] \ ,
\end{equation}
where have explicitly kept the one-loop part and $g \geq 2$ corresponds
to higher loops.
Solving for $u_0$ in this expansion, and substituting in the
loop-corrected entropy
one finds the form
\begin{equation}
 S(T) = \pi^3 b N \hat{V}_5 \frac{\thg}{\thg - T} \left[
1 + \sum_{n=1}^\infty k_{n} (N) \left( \frac{\thg-T}{\thg}
\right)^{n} \right] \ ,
\end{equation}
in agreement with the higher order terms \eqref{Sho}, obtained from the statistical thermodynamics.

\section{Conclusions}

In this letter we have successfully implemented string corrections
in the near-horizon NS5-brane thermodynamics to explain the
Hagedorn behaviour of LST via the near-horizon-NS5/LST duality.
The tree-level corrections to the temperature were seen to be part
of the Hagedorn temperature, and it was explained that these
corrections should vanish in order for the correspondence to work.
To support this, it was argued that the string-frame metric should
not have tree-level corrections. Subsequently, the leading
one-loop string correction was shown to exactly generate the type
of corrections to the temperature necessary to reproduce the
correct temperature dependence of the entropy of LST near the
Hagedorn temperature. The one-loop correction of the temperature
was also argued to imply that the Hagedorn temperature of LST is a
limiting temperature which cannot be crossed. Moreover,
higher-loop corrections were seen to be in agreement with the
expected form in this framework.

The corrected near-horizon NS5-brane thermodynamics gave a prediction
in \eqref{k} for the constant $k$ which characterizes
the single string density of levels \eqref{sing0}.
At present it is unclear how to reproduce
this constant from the LST side, and it would be very interesting to
obtain a better understanding of this. In critical superstring
theory one has instead $k=1$ \cite{Brandenberger:1989aj}.

We note that an analysis similar to that of this letter can be
done for the heterotic five-brane. In that case one still
finds that the entropy is proportional to $(\thg -T)^{-1}$ but with
a different factor $k$. The dual theory is a LST with 8 supercharges.

Finally, we report on a preliminary result in the case of NCLST
with 16 supercharges. NCLST is LST on a non-commutative space,
which has a dual description in terms of delocalized D2-branes or
D3-branes for type IIA and type IIB string theory respectively
\cite{Alishahiha:2000er} \footnote{This is related to the
correspondence between non-commutative theories and near-horizon
brane configurations \cite{Maldacena:1999mh,Lu:1999rm}.}. 
The entropy in
this case should be proportional to \( (\thg - T)^{-2/3} \)
\cite{Correia:2000}, a behaviour that is very different from that
of LST. It would be interesting if one could obtain this behaviour
from the statistical mechanics of closed strings on a
non-commutative space.

\section*{Acknowledgments}

We thank J. Correia, P. Di Vecchia, E. Kiritsis, F. Larsen, A.
Liccardo, J. L. Petersen, A. Sevrin and especially J. de Boer for
useful discussions. This work is supported in part by TMR network
ERBFMRXCT96-0045.

\begin{appendix}

\renewcommand{\theequation}{A.\arabic{equation}}
\section{Hagedorn behaviour in string theory}

The statistical mechanics of string theory breaks down at sufficiently
high temperatures
\cite{Hagedorn:1965st,Salomonson:1986eq,Atick:1988si,Brandenberger:1989aj,Bowick:1989us,Lowe:1995nm}.
At the so-called Hagedorn temperature \( \thg \) the partition
function is ill-defined since the density of states in string theory
grows exponentially.

For a closed string theory in a $D$-dimensional space-time with
central charge $c$, the number of states
$d_L(n_L)$ and $d_R(n_R)$ in the left and right
moving sectors respectively, are given by the asymptotic formulae
\cite{Green:1987sp}
\begin{equation}
\label{dns} d_L(n_L) \propto n_L^{-\frac{D+1}{4}} \exp(
\bhg\sqrt{n_L}/l_s ) \sp d_R(n_R) \propto n_R^{-\frac{D+1}{4}}
\exp( \bhg \sqrt{n_R}/l_s ) \ ,
\end{equation}
where $\bhg=\thg^{-1}$ is the inverse Hagedorn temperature,
given by
\begin{equation}
\label{canthg} \thg = \frac{1}{2\pi l_s} \sqrt{\frac{6}{c}} \ .
\end{equation}
Here, the central charge is $c=\frac{3}{2}(D-2)$ for supersymmetric
string theory and $c=D-2$ for bosonic string theory.

The single string density of states for free closed strings
in $d$ non-compact space directions is then \cite{Salomonson:1986eq,Brandenberger:1989aj,Bowick:1989us,Lowe:1995nm}
\begin{equation}
\label{sing} \omega (E) \propto E^{-\frac{d}{2}-1} \exp ( \bhg E )
\ .
\end{equation}
The Canonical partition function for a gas of weakly interacting closed
strings is
\begin{equation}
\label{partsum} Z(\beta) = \sum_{n=1}^\infty \frac{1}{n!}
\int_{\Lambda}^\infty \prod_{i=1}^n dE_i \ \omega(E_i) \exp(
-\beta E_i ) \ ,
\end{equation}
where $\Lambda$ is the IR-cutoff. This gives
\begin{equation}
\label{logpart} \log( 1 + Z(\beta) ) = \int_{\Lambda}^\infty dE \
\omega (E) \exp(-\beta E) \ .
\end{equation}
For \( d=0 \) we have the single-string density of states
\begin{equation}
\label{sing0} \omega (E) = k  E^{-1} \exp ( \bhg E ) \ .
\end{equation}
Using this in \eqref{logpart}, the mean energy is given by
\begin{equation}
\label{ET}
E = - \frac{\partial}{\partial \beta} \log(Z(\beta)) =
\frac{k}{\beta-\bhg} \ ,
\end{equation}
which gives the following entropy
\begin{equation}
\label{hagentr} S(E) = \bhg E + k \log E \ .
\end{equation}
We note that in the Microcanonical ensemble we have $ S(E) = \bhg
E + (k-1) \log E $ instead \cite{Brandenberger:1989aj}. This means
that the Canonical and Microcanonical ensembles are not equivalent
in general \cite{Brandenberger:1989aj,Lowe:1995nm}, but since in
our application in the main text \( k \gg 1 \), the ensembles are
effectively equivalent in this case.

From \eqref{hagentr} we find the entropy as a function of temperature
\begin{equation}
\label{hagentrT} S(T) = k \frac{\thg}{\thg- T} \ . 
\end{equation}
Expanding \eqref{logpart}, the entropy including higher order terms reads
\begin{equation}
\label{Sho} S(T) = k \frac{\thg}{\thg - T} \left[ 1 +
\sum_{n=1}^\infty k_n \left( \frac{\thg-T}{\thg} \right)^n \right]
\ .
\end{equation}

\end{appendix}


\addcontentsline{toc}{section}{References}

\begin{thebibliography}{10}

\bibitem{Aharony:1998ub}
O.~Aharony, M.~Berkooz, D.~Kutasov, and N.~Seiberg, {\it Linear dilatons,
  {NS5-branes} and holography},  {\em JHEP} {\bf 10} (1998) 004,
  [\href{http://xxx.lanl.gov/abs/hep-th/9808149}{{\tt hep-th/9808149}}].

\bibitem{Seiberg:1997zk}
N.~Seiberg, {\it New theories in six-dimensions and matrix description of {M}
  theory on {$T^5$} and {$T^5/Z_2$}},  {\em Phys. Lett.} {\bf B408} (1997)
  98--104, [\href{http://xxx.lanl.gov/abs/hep-th/9705221}{{\tt
  hep-th/9705221}}];
M.~Berkooz, M.~Rozali, and N.~Seiberg, {\it Matrix description of {M} theory on
  {$T^4$} and {$T^5$}},  {\em Phys. Lett.} {\bf B408} (1997) 105--110,
  [\href{http://xxx.lanl.gov/abs/hep-th/9704089}{{\tt hep-th/9704089}}].

\bibitem{Dijkgraaf:1997ku}
R.~Dijkgraaf, E.~Verlinde, and H.~Verlinde, {\it Notes on matrix
and micro
  strings},  {\em Nucl. Phys. Proc. Suppl.} {\bf 62} (1998) 348,
  [\href{http://xxx.lanl.gov/abs/hep-th/9709107}{{\tt hep-th/9709107}}].

\bibitem{Losev:1997hx}
A.~Losev, G.~Moore, and S.~L. Shatashvili, {\it M {\&} m's},  {\em Nucl. Phys.}
  {\bf B522} (1998) 105--124,
  [\href{http://xxx.lanl.gov/abs/hep-th/9707250}{{\tt hep-th/9707250}}].

\bibitem{Aharony:1999ks}
O.~Aharony, {\it A brief review of 'little string theories'},  {\em Class.
  Quant. Grav.} {\bf 17} (2000) 929,
  [\href{http://xxx.lanl.gov/abs/hep-th/9911147}{{\tt hep-th/9911147}}].

\bibitem{Maldacena:1996ya}
J.~M. Maldacena, {\it Statistical entropy of near extremal five-branes},  {\em
  Nucl. Phys.} {\bf B477} (1996) 168--174,
  [\href{http://xxx.lanl.gov/abs/hep-th/9605016}{{\tt hep-th/9605016}}].

\bibitem{Maldacena:1997cg}
J.~M. Maldacena and A.~Strominger, {\it Semiclassical decay of near-extremal
  fivebranes},  {\em JHEP} {\bf 12} (1997) 008,
  [\href{http://xxx.lanl.gov/abs/hep-th/9710014}{{\tt hep-th/9710014}}].

\bibitem{Harmark:1999xt}
T.~Harmark and N.~A. Obers, {\it Thermodynamics of spinning branes and their
  dual field theories},  {\em JHEP} {\bf 01} (2000) 008,
  [\href{http://xxx.lanl.gov/abs/hep-th/9910036}{{\tt hep-th/9910036}}];
T.~Harmark and N.~A. Obers, {\it Thermodynamics of field theories from spinning
  branes},  \href{http://xxx.lanl.gov/abs/hep-th/0002250}{{\tt
  hep-th/0002250}}.

\bibitem{Kiritsis:1999ke}
E.~Kiritsis and T.~R. Taylor, {\it Thermodynamics of {D-brane}
probes}, \href{http://xxx.lanl.gov/abs/hep-th/9906048}{{\tt
  hep-th/9906048}}. 

\bibitem{Strominger:1996sh}
A.~Strominger and C.~Vafa, {\it Microscopic origin of the {Bekenstein-Hawking}
  entropy},  {\em Phys. Lett.} {\bf B379} (1996) 99--104,
  [\href{http://xxx.lanl.gov/abs/hep-th/9601029}{{\tt hep-th/9601029}}];
C.~G. {Callan Jr.} and J.~M. Maldacena, {\it D-brane approach to black hole
  quantum mechanics},  {\em Nucl. Phys.} {\bf B472} (1996) 591--610,
  [\href{http://xxx.lanl.gov/abs/hep-th/9602043}{{\tt hep-th/9602043}}];
G.~T. Horowitz and A.~Strominger, {\it Counting states of near-extremal black
  holes},  {\em Phys. Rev. Lett.} {\bf 77} (1996) 2368--2371,
  [\href{http://xxx.lanl.gov/abs/hep-th/9602051}{{\tt hep-th/9602051}}].

\bibitem{Diaconescu:1997gu}
D.-E. Diaconescu and N.~Seiberg, {\it The {Coulomb} branch of (4,4)
  supersymmetric field theories in two dimensions},  {\em JHEP} {\bf 07} (1997)
  001, [\href{http://xxx.lanl.gov/abs/hep-th/9707158}{{\tt hep-th/9707158}}].

\bibitem{Maldacena:1997re}
J.~Maldacena, {\it The large {N} limit of superconformal field theories and
  supergravity},  {\em Adv. Theor. Math. Phys.} {\bf 2} (1998) 231--252,
  [\href{http://xxx.lanl.gov/abs/hep-th/9711200}{{\tt hep-th/9711200}}];
O.~Aharony, S.~S. Gubser, J.~Maldacena, H.~Ooguri, and Y.~Oz, {\it Large {N}
 field theories, string theory and gravity},  {\em Phys. Rept.} {\bf 323}
  (2000) 183, [\href{http://xxx.lanl.gov/abs/hep-th/9905111}{{\tt
  hep-th/9905111}}].    

\bibitem{Sfetsos:1999pq}
K.~Sfetsos, {\it Rotating {NS5}-brane solution and its exact string theoretical
  description},  {\em Fortsch. Phys.} {\bf 48} (2000) 199--204,
  [\href{http://xxx.lanl.gov/abs/hep-th/9903201}{{\tt hep-th/9903201}}].    

\bibitem{Ooguri:1996wj}
H.~Ooguri and C.~Vafa, {\it Two-dimensional black hole and singularities of
  {CY} manifolds},  {\em Nucl. Phys.} {\bf B463} (1996) 55--72,
  [\href{http://xxx.lanl.gov/abs/hep-th/9511164}{{\tt hep-th/9511164}}].

\bibitem{Douglas:1996sw}
M.~R. Douglas and G.~Moore, {\it {D-branes}, quivers, and {ALE} instantons},
  \href{http://xxx.lanl.gov/abs/hep-th/9603167}{{\tt hep-th/9603167}};
C.~V. Johnson and R.~C. Myers, {\it Aspects of type {IIB} theory on {ALE}
  spaces},  {\em Phys. Rev.} {\bf D55} (1997) 6382--6393,
  [\href{http://xxx.lanl.gov/abs/hep-th/9610140}{{\tt hep-th/9610140}}].

\bibitem{Hashimoto:1996pd}
A.~Hashimoto, {\it Perturbative dynamics of fractional strings on multiply
  wound {D-strings}},  {\em Int. J. Mod. Phys.} {\bf A13} (1998) 903,
  [\href{http://xxx.lanl.gov/abs/hep-th/9610250}{{\tt hep-th/9610250}}].

\bibitem{Correia:2000}
J.~Correia, T.~Harmark, and N.~A. Obers. In preparation.
                                                                
\bibitem{Gross:1986iv}
D.~J. Gross and E.~Witten, {\it Superstring modifications of {Einstein's}
  equations},  {\em Nucl. Phys.} {\bf B277} (1986) 1.

\bibitem{Green:1997tv}
M.~B. Green and M.~Gutperle, {\it Effects of {D-instantons}},  {\em Nucl.
  Phys.} {\bf B498} (1997) 195,
  [\href{http://xxx.lanl.gov/abs/hep-th/9701093}{{\tt hep-th/9701093}}].

\bibitem{Callan:1991dj}
C.~G. {Callan Jr.}, J.~A. Harvey, and A.~Strominger, {\it World sheet approach
  to heterotic instantons and solitons},  {\em Nucl. Phys.} {\bf B359} (1991)
  611;
C.~G. {Callan Jr.}, J.~A. Harvey, and A.~Strominger, {\it Worldbrane actions
  for string solitons},  {\em Nucl. Phys.} {\bf B367} (1991) 60--82.

\bibitem{Sfetsos:1998wc}
K.~Sfetsos, {\it On (multi-)center branes and exact string vacua},
  \href{http://xxx.lanl.gov/abs/hep-th/9812165}{{\tt hep-th/9812165}}.

\bibitem{Tseytlin:1993df}
A.~A. Tseytlin, {\it On field redefinitions and exact solutions in string
  theory},  {\em Phys. Lett.} {\bf B317} (1993) 559--564,
  [\href{http://xxx.lanl.gov/abs/hep-th/9308042}{{\tt hep-th/9308042}}].

\bibitem{Salomonson:1986eq}
P.~Salomonson and B.-S. Skagerstam, {\it On superdense superstring gases: A
  heretic string model approach},  {\em Nucl. Phys.} {\bf B268} (1986) 349.

\bibitem{Lowe:1995nm}
D.~A. Lowe and L.~Thorlacius, {\it Hot string soup},  {\em Phys. Rev.} {\bf
  D51} (1995) 665--670, [\href{http://xxx.lanl.gov/abs/hep-th/9408134}{{\tt
  hep-th/9408134}}].

\bibitem{Giveon:1999px}
A.~Giveon and D.~Kutasov, {\it Little string theory in a double scaling limit},
   {\em JHEP} {\bf 10} (1999) 034,
  [\href{http://xxx.lanl.gov/abs/hep-th/9909110}{{\tt hep-th/9909110}}]; 
A.~Giveon and D.~Kutasov, {\it Comments on double scaled little string theory},
   {\em JHEP} {\bf 01} (2000) 023,
  [\href{http://xxx.lanl.gov/abs/hep-th/9911039}{{\tt hep-th/9911039}}].

\bibitem{Brandenberger:1989aj}
R.~Brandenberger and C.~Vafa, {\it Superstrings in the early universe},  {\em
  Nucl. Phys.} {\bf B316} (1989) 391.

\bibitem{Barbon:1998cr}
J.~L.~F. Barbon, I.~I. Kogan, and E.~Rabinovici, {\it On stringy thresholds in
  {SYM}/{AdS} thermodynamics},  {\em Nucl. Phys.} {\bf B544} (1999) 104,
  [\href{http://xxx.lanl.gov/abs/hep-th/9809033}{{\tt hep-th/9809033}}]; 
S.~A. Abel, J.~L.~F. Barbon, I.~I. Kogan, and E.~Rabinovici, {\it String
  thermodynamics in {D-brane} backgrounds},  {\em JHEP} {\bf 04} (1999) 015,
  [\href{http://xxx.lanl.gov/abs/hep-th/9902058}{{\tt hep-th/9902058}}]; 
I.~Antoniadis, J.~P. Derendinger, and C.~Kounnas, {\it Non-perturbative
  temperature instabilities in {N=4} strings},  {\em Nucl. Phys.} {\bf B551}
  (1999) 41, [\href{http://xxx.lanl.gov/abs/hep-th/9902032}{{\tt
  hep-th/9902032}}].                                                   

\bibitem{Alishahiha:2000er}
M.~Alishahiha, {\it On type {II} {NS5-branes} in the presence of an {RR}
  field},  \href{http://xxx.lanl.gov/abs/hep-th/0002198}{{\tt hep-th/0002198}}.

\bibitem{Maldacena:1999mh}
J.~M. Maldacena and J.~G. Russo, {\it Large {N} limit of non-commutative gauge
  theories},  {\em JHEP} {\bf 09} (1999) 025,
  [\href{http://xxx.lanl.gov/abs/hep-th/9908134}{{\tt hep-th/9908134}}];
A.~Hashimoto and N.~Itzhaki, {\it Non-commutative {Yang--Mills} and the
  {AdS/CFT} correspondence},  {\em Phys. Lett.} {\bf B465} (1999) 142,
  [\href{http://xxx.lanl.gov/abs/hep-th/9907166}{{\tt hep-th/9907166}}];
M.~Li and Y.-S. Wu, {\it Holography and noncommutative {Yang--Mills}},
{\em Phys. Rev. Lett.} {\bf 84} (2000) 2084,
  [\href{http://xxx.lanl.gov/abs/hep-th/9909085}{{\tt hep-th/9909085}}];       
M.~Alishahiha, Y.~Oz, and M.~M. Sheikh-Jabbari, {\it Supergravity and large {N}
  noncommutative field theories},  {\em JHEP} {\bf 11} (1999) 007,
  [\href{http://xxx.lanl.gov/abs/hep-th/9909215}{{\tt hep-th/9909215}}];
T.~Harmark and N.~A. Obers, {\it Phase structure of non-commutative field
  theories and spinning brane bound states},  {\em JHEP} {\bf 03} (2000) 024,
  [\href{http://xxx.lanl.gov/abs/hep-th/9911169}{{\tt hep-th/9911169}}].

\bibitem{Lu:1999rm}
J.~X. Lu and S.~Roy, {\it {$(p+1)$}-dimensional noncommutative {Yang--Mills}
  and {D$(p-2)$} branes},  \href{http://xxx.lanl.gov/abs/hep-th/9912165}{{\tt
  hep-th/9912165}}; 
R.-G. Cai and N.~Ohta, {\it Noncommutative and ordinary super {Yang--Mills} on
{(D$(p-2)$, D$p$)} bound states},  {\em JHEP} {\bf 03} (2000) 009,
  [\href{http://xxx.lanl.gov/abs/hep-th/0001213}{{\tt hep-th/0001213}}].    

\bibitem{Hagedorn:1965st}
R.~Hagedorn, {\it Statistical thermodynamics of strong interactions at high-
  energies},  {\em Nuovo Cim. Suppl.} {\bf 3} (1965) 147--186.

\bibitem{Atick:1988si}
J.~J. Atick and E.~Witten, {\it The {Hagedorn} transition and the number of
  degrees of freedom of string theory},  {\em Nucl. Phys.} {\bf B310} (1988)
  291.

\bibitem{Bowick:1989us}
M.~J. Bowick and S.~B. Giddings, {\it High temperature strings},  {\em Nucl.
  Phys.} {\bf B325} (1989) 631.

\bibitem{Green:1987sp}
M.~B. Green, J.~H. Schwarz, and E.~Witten, {\em Superstring theory. Vol. 1:
  Introduction}.
\newblock Cambridge monographs on mathematical physics. Cambridge University
  Press, 1987.

\end{thebibliography}
\providecommand{\href}[2]{#2}\begingroup\raggedright
 \endgroup
\end{document}